\documentclass[aps,prb,twocolumn,superscriptaddress,amsmath,amssymb,showpacs,superscriptaddress]{revtex4-1}

\usepackage{graphicx}
\usepackage{dcolumn}
\usepackage{bm}
\usepackage{hyperref}
\usepackage{color}

\begin{document}


\title{Poisson noise induced switching in driven micromechanical resonators}

\author{J. Zou}
\author{S. Buvaev}
\affiliation{Department of Physics, University of Florida, Gainesville, Florida 32611, USA }
\author{M. Dykman}
\affiliation{Department of Physics and Astronomy, Michigan State University, East Lansing, Michigan 48824, USA }
\author{H. B. Chan}
\email[]{hochan@ust.hk}
\affiliation{Department of Physics, the Hong Kong University of Science and Technology, Hong Kong, China }

\date{\today}

\begin{abstract}

We study Poisson-noise induced switching between coexisting vibrational states in driven nonlinear micromechanical resonators. In contrast to Gaussian noise induced switching, the measured logarithm of the switching rate is proportional not to the reciprocal noise intensity, but to its logarithm, for fixed pulse area. We also find that the switching rate logarithm varies as a square root of the distance to the bifurcation point, instead of the conventional scaling with exponent 3/2.

\end{abstract}


\maketitle
\section{INTRODUCTION}
Non-Gaussian fluctuations in mesoscopic devices and in photon statistics have proved to be a powerful probe of the underlying physics of such systems, yielding information that cannot be deduced from the averaged value and the variance.\cite{Levitov1996} Experimentally, detecting a deviation from the Gaussian distribution is a challenging task. Direct measurements of the third and higher moments of fluctuations in mesoscopic systems require highly sophisticated techniques.\cite{Reulet2003,Bomze2005,Gustavsson2006,Fricke2010} In another approach, following theoretical proposals,\cite{Tobiska2004,Pekola2004,Jordan2005,*Sukhorukov2007,Grabert2008} non-Gaussian statistics was detected using noise-activated switching out of a metastable state in a Josephson junction.\cite{Timofeev2007,LeMasne2009} This approach provides direct access to the tail of the noise distribution, as a relatively large noise outburst is required to drive the Josephson junction out of the effective potential well. In the measurements, deviations from the Gaussian statistics were found from the asymmetry of the switching rate with respect to the current polarity.

Noise-induced switching has attracted much attention recently in the context of nonequilibrium systems, where the notion of escape from a potential well does not apply. In particular, a number of experiments were done on periodically driven systems, with the metastable states corresponding to forced vibrations. The systems ranged from trapped electrons\cite{Lapidus1999} to Josephson junctions,\cite{Siddiqi2006,*Vijay2009} mechanical resonators, \cite{Aldridge2005,Stambaugh2006b,*Chan2007a} and trapped atoms.\cite{Kim2006} It was found that, for Gaussian noise, the switching rate $W$ displays universal features. Its dependence on the noise intensity $D_G$ is of the activation form,  $W \propto\exp(-{\cal Q})$, with the switching exponent ${\cal Q} = R_G/D_G$, where $R_G$ is the effective activation barrier. Furthermore, $R_G$ displays the predicted  \cite{Dykman1979a,*Dykman1980} scaling behavior close to bifurcation points where the number of stable states changes,\cite{Aldridge2005,Siddiqi2006,*Vijay2009,Stambaugh2006b,*Chan2007a} $R_G\propto\eta^\xi$, where $\eta$ is the distance to the bifurcation point.

In this paper we use a driven micromechanical oscillator to study the features of Poisson-noise induced switching. In contrast to Gaussian noise, Poisson noise is characterized not just by its intensity $D_P$ and correlation time, but also by the pulse rate $\nu_P$. We limit ourselves to the most interesting case of short pulses and explore the dependence of the switching rate on the pulse area and $\nu_P$ as well as the distance to the bifurcation point $\eta$. Even though the stable states of this nonequilibrium system are not separated by a potential barrier, our findings are consistent with the recently predicted universal behavior of Poisson-noise induced switching near a bifurcation point, which differs qualitatively from that for Gaussian noise \cite{Billings2010,*Dykman2010a}.

\begin{figure}[h]
\includegraphics[width=6.5cm]{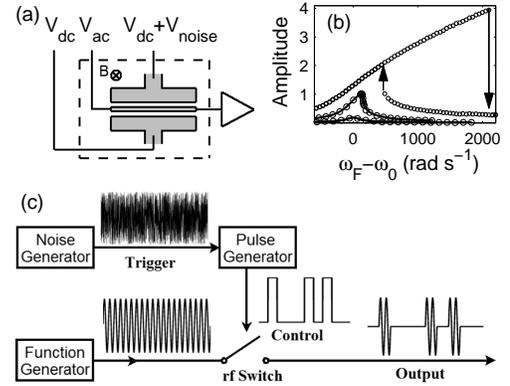}
\caption{\label{fig:Fig1} (a) Schematic of the experimental setup. The noise force is generated by applying voltage $V_{\textrm{dc}} + V_{\textrm{noise}}$ to the top sidegate, with $|V_{\textrm{dc}}|\gg |V_{\textrm{noise}}|$. (b) Vibration amplitude vs. driving frequency; the vibration amplitude  is scaled by its value at the critical point where hysteresis emerges. The curves correspond to the driving amplitudes of $0.2A_c$, $A_c$, and $4A_c$, where $A_c$ is the critical driving amplitude for the onset of hysteresis. The up-arrow indicates the position of the bifurcation point $\omega_1$. We fit the linear and the critical responses to extract the parameters of the oscillator. (c) Schematic of the generation of Poisson pulses. Each output pulse contains $>$ 400 rf cycles. For clarity, only 2 rf cycles are shown.}
\label{fig1}
\end{figure}

\section{EXPERIMENTAL SETUP}

Our oscillator is a double-clamped polysilicon beam that is $100~\mu$m long and $1.2~\mu$m by $1.5~\mu$m in cross section. By passing an ac current through the beam in a perpendicular magnetic field of 5 T, we excited motion of the beam in its in-plane fundamental mode [Fig. 1(a)]. This motion generates an electromotive force that changes the transmitted ac power. All measurements were performed at $4$~K and $< 10^{-6}$~torr. The mode can be modeled by a Duffing oscillator \cite{Aldridge2005,Kozinsky2007,Unterreithmeier2010},
\begin{equation}
\frac{d^2q}{dt^2}+2\Gamma \frac{dq}{dt}+\omega_0^2 q +\gamma q^3 = A \cos \omega_F t + f(t),
\label{eq:Duffing}
\end{equation}
where $q$ is the normalized beam displacement, $\Gamma=96$ rad$~$s$^{-1}$ is the damping coefficient,  $\omega_0 = 7,133,339$ rad$~$s$^{-1}$ is the resonant frequency, $\gamma = 2.3\times 10^9~\textrm{s}^{-2}$ is the coefficient of the cubic nonlinearity, $A$ and $\omega_F$ are, respectively, the amplitude and frequency of the external driving force, and $f(t)$ is the noise.

In the absence of noise, for small $A$ the device responds as a weakly damped harmonic oscillator [Fig.~\ref{fig1}~(b)]. As $A$ increases, the response curve bends toward higher frequencies because of spring hardening. For $A$ exceeding the critical value $A_c$, the oscillator becomes bistable, with two coexisting states of forced vibrations in the range $\omega_1<\omega_F < \omega_2$. The frequencies $\omega_{1,2}$ depend on $A$. They correspond to saddle-node bifurcations and are indicated by arrows in Fig.~\ref{fig1}~(b). 

Noise can lead to switching between the vibrational states. In the parameter range of our operation, thermal noise was too weak for switching to be observed. We generated noise externally, by applying random electrostatic forces between one of the sidegates and the beam. To create Gaussian noise, $f(t)=f_G(t)$ in Eq.~(\ref{eq:Duffing}), the Johnson noise from a $50~\Omega$ resistor at room temperature was amplified and directly applied to the sidegate.

Generating Poisson noise involves extra circuitry [Fig.~\ref{fig1}~(c)]. Since the oscillator is most perceptive to noise at frequencies close to $\omega_0$,  we used noise that consisted of pulses at frequency $\omega_F$ with duration $t_P\gg \omega_F^{-1}$. In the rotating frame, as we will explain later, such pulses looked rectangular. To create them, the Gaussian noise voltage was connected to the trigger input of a pulse generator. Whenever the input voltage, in a rare large outburst, exceeded a threshold value, the generator produced a square pulse. The pulse width $t_P =400~\mu$s was much less than the reciprocal pulse rate $\nu_P^{-1}$, which ranged from $30$~ms to $200$~ms. We verified that the inter-pulse intervals obeyed the Poisson statistics. The pulse signal controlled an rf switch that only turned on when a pulse was present, generating an electrostatic force with the same phase as the regular magnetomotive force. The output of the switch is described by the random force in Eq.~(\ref{eq:Duffing}) of the form
\begin{equation}
\label{eq:Poisson_noise}
f(t)\equiv f_P(t)= \bar f_P(t)\cos\omega_Ft,
\end{equation}
where $\bar f_P(t)$ is a random pulse train, with the area of each pulse $g_P$ controlled by a variable attenuator; for chosen $t_P$ there were $\approx 454$~rf cycles in each pulse. The pulses represent Poisson noise with intensity $D_P=\nu_Pg_P^2/2$. We checked that a single pulse never led to switching.

\section{Theoretical background}
\label{sec:theory}

To study the oscillator dynamics, we go to the rotating frame and introduce slow variables $X$ and $Y$,
%
$q(t)=C_{\textrm{res}}(X\cos \omega_F t + Y\sin \omega_F t)$,
$\dot{q}(t)=-\omega_F C_{\textrm{res}}(X\sin \omega_F t - Y\cos \omega_F t)$,
%
where $C_{\textrm{res}}=\left(8\omega_F\delta\omega/3\gamma\right)^{1/2}$ and $\delta\omega = \omega_F-\omega_0$ is the frequency detuning. Functions $X$ and $Y$ are the scaled in-phase and quadrature components of $q(t)$. In the rotating wave approximation (RWA)  Poisson noise (\ref{eq:Poisson_noise}) is demodulated and becomes a random force $\tilde f_P$ that drives the quadrature component $Y$. In dimensionless ``slow" time $\tau =t\delta\omega \equiv t(\omega_F-\omega_0)$  Eq.~(\ref{eq:Duffing}) becomes
\begin{eqnarray}
\label{eq:eom_full}
&&\dot X= K_X, \qquad \dot Y=K_Y + \tilde f_P(\tau),
\nonumber\\
&&K_X=-\Omega^{-1}X+\partial_Y {\cal G},\qquad K_Y=-\Omega^{-1}Y -\partial_X{\cal G}, \nonumber \\
&&{\cal G}(X,Y)=\frac{1}{4}(X^2+Y^2-1)^2- \beta^{1/2}X.
\end{eqnarray}
Here, dot indicates differentiation with respect to $\tau$, and the noise $\tilde f_P(\tau)=\bar f_P(t)/2\omega_FC_{\rm res}\delta\omega$. The two dimensionless parameters $\beta$ and $\Omega$ are the scaled intensity of the modulation and the scaled detuning of the modulation frequency from the oscillator eigenfrequency,
\begin{equation}
\label{eq:beta_Omega}
\beta= 3\gamma A^2/32\omega_F^3(\delta\omega)^3,\qquad\Omega^{-1}=\Gamma/\delta\omega.
\end{equation}

Functions $K_X,K_Y$ do not explicitly depend on time in the rotating frame. The stationary solutions of equations (\ref{eq:eom_full}) in the absence of noise, which are given by equations $K_X=K_Y=0$, determine the stationary vibrational states of the oscillator. In the region of bistability there are three stationary states, two of them are stable and the third is the saddle-type solution of Eq.~(\ref{eq:eom_full}) in the absence of noise.

For chosen pulse duration $t_P\ll \Gamma^{-1}$, the noise $\tilde f_P(\tau)$ is effectively $\delta$-correlated.  From Eqs.~(\ref{eq:Duffing}) and (\ref{eq:Poisson_noise}), each pulse of $\tilde f_P(\tau)$ shifts $Y(\tau)$ by
$\tilde g_P=(2\omega_FC_{\rm res})^{-1}g_P$.  The dimensionless pulse rate is $\tilde\nu_P=\nu_P/\delta\omega$ and the dimensionless intensity is $\tilde D_P=\nu_Pg_P^2/8\omega_F^2C_{\rm res}^2\delta\omega$.

If the system is initially at a stable state, for small $\nu_P/\Gamma$ and $\tilde g_P$ it usually has time to relax back to this state between the pulses. However, occasionally there may happen a sequence of frequent pulses that will push the system far from the occupied state so as to cause switching to a different state. This switching mechanism is qualitatively different from the case of Gaussian noise where individual noise pulses are too frequent to be resolved by the system and the noise outburst has to overcome the deterministic force that drives the system toward the stable state.

It is most interesting to study Poisson-noise induced switching close to the saddle-node bifurcation point where, on the one hand, the switching rate $W$ becomes larger and, on the other hand, the system dynamics is expected to display universal features related to the critical slowing down. The quadrature $Y(\tau)$ corresponds to the soft mode near a bifurcation point of the Duffing oscillator. The in-phase component $X(\tau)$ follows $Y(\tau)$ adiabatically, so that  \cite{Dykman1979a}
\begin{equation}
\label{eq:a_B}
X(\tau)-X_B\approx a_B[Y(\tau)-Y_B],\qquad a_B=\Omega(2r_B^2-1),
\end{equation}
where $X_B$ and $Y_B$ are the values of $X$ and $Y$ at the bifurcation point and $r_B^2=X_B^2+Y_B^2$.

Extending the analysis \cite{Dykman1979a} to the Poisson noise, one obtains equation of motion for the soft mode as $\dot Y=-\partial U/\partial Y + \tilde f_P(\tau)$. The potential $U$ has the form  $U(Y)=c_1y^3 + c_2\eta y$, where $y=Y-Y_B$ is the distance to the bifurcational value  of $Y$. The parameters $c_{1,2}$ in the expression for $U$ are known;\cite{Dykman1979a} $\eta$ is the distance to the bifurcation point in the space of the driving force parameters $(A,\omega_F)$; If $c_1c_2\eta <0$,  potential $U(Y)$ has a minimum and a maximum at $Y_{\min}$ and $Y_{\max}$, respectively. The minimum corresponds to the stable state of forced vibrations. The extrema of $U(y)$ merge at the bifurcation point $\eta =0$ and disappear for $c_1c_2\eta > 0$. This maps the problem of the oscillator dynamics near a bifurcation point onto the problem of an overdamped particle with one dynamical variable in a static potential.

Using the results for the switching rate  $W\propto \exp(-{\cal Q})$ near a bifurcation point in a static potential,\cite{Billings2010}, we obtain for the switching exponent ${\cal Q}={\cal Q}_P$
\begin{eqnarray}
\label{eq:QP_analytic}
\mathcal{Q}_P \approx (2\eta^{1/2}/g)\log(\kappa\eta/g\nu), \qquad \eta = \beta_B - \beta.
\end{eqnarray}
Here
\begin{equation}
\label{eq:define_nu}
g=\frac{(2|b|)^{1/2}\beta_B^{1/4}}{2\omega_FC_{B}}g_P,\qquad \nu=\frac{(2/|b|)^{1/2}\beta_B^{1/4}}{\delta\omega_B}\nu_P,
\end{equation}
with $b=\beta_B^{1/2}(\delta\omega_B)^2(3r_B^2-2)/2\Gamma^2$. Parameters $\beta_B$, $\delta\omega_B$, and $C_B$ are the bifurcational values of the scaled modulation intensity $\beta$ given by Eq.~(\ref{eq:beta_Omega}), the frequency detuning $\delta\omega$, and the scaling factor $C_{\rm res}$, respectively; parameter $\kappa$ in Eq.~(\ref{eq:QP_analytic}) is $\sim 1$. Equation (\ref{eq:QP_analytic}) applies provided the number of noise pulses $n_{\rm sw}$ needed to shift the system from $Y_{\min}$ to $Y_{\max}$ is $n_{\rm sw}\gg 1,(\nu_P/\delta\omega)\tau_{\rm r}$, where the dimensionless relaxation time is $\tau_{\rm r}=1/U''(Y_{\min})$. We checked that this condition was met in the experiment. Intuitively, ${\cal Q}_P$ can be found by maximizing the probability for the Poisson distribution to have $n_{\rm sw}\sim (Y_{\max}-Y_{\min})/\tilde g_P \propto \eta^{1/2}/g$ pulses in time $\tau_{\rm r}$ and taking the logarithm of this probability.\cite{Billings2010}

Equation (\ref{eq:QP_analytic}) should be contrasted with the expression for the switching exponent ${\cal Q} \equiv {\cal Q}_G$ in the case of Gaussian noise, $f(t)=f_G(t)$ in Eq.~(\ref{eq:Duffing}). For white noise, $\langle f_G(t)f_G(t')\rangle = 2D_G\delta(t-t')$, near a bifurcation point\cite{Dykman1980}
\begin{equation}
\mathcal{Q}_G \approx \frac{4|\eta|^{3/2}}{3\tilde D_G},\qquad
\tilde D_G=D_G\frac{\beta_B^{3/4}|2b|^{1/2}}{\omega_F^2C_{B}^2\delta\omega_B}.
\label{eq:QG_analytic}
\end{equation}
This expression displays a different dependence on $\eta$ and the noise parameters than Eq.~(\ref{eq:QP_analytic}).

\section{RESULTS AND DISCUSSION}

\begin{figure}[h]
\includegraphics[width=6.8cm]{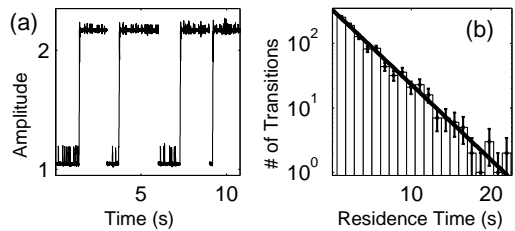}
\caption{(a) Transitions from the low-amplitude state to the high-amplitude state at $\Delta\omega_{\eta}=9.4$~rad$~$s$^{-1}$ in the presence of Poisson noise. The amplitude is normalized to the vibration amplitude at the critical point. The small spikes are the response of the oscillator to pulses that do not lead to transitions. The response during reset to the low-amplitude state is not shown. (b) The histogram of the residence times. The linear fit gives the mean residence time. }
\label{fig2}
\end{figure}

In the experiment, the driving amplitude $A$ was $\approx 4 A_c$, so that the hysteresis loop in Fig.~\ref{fig1}~(b) was comparatively large while the vibrations remained almost sinusoidal. We measured the switching rate near the bifurcation point $\omega_{1B}$ at driving frequency
\[\omega_F=\omega_{1B}+\Delta\omega_{\eta},\qquad \Delta\omega_{\eta}\ll|\omega_{1B}-\omega_0|;\]
the distance to the bifurcation point was along the frequency axis, with $\Delta\omega_\eta \propto \eta$.

The oscillator was initially prepared in the low-amplitude state. Then the noise was turned on, and we monitored the oscillator amplitude. Figure~\ref{fig2}~(a) exhibits a time series of the measured amplitude in the presence of Poisson noise.  After switching to the high amplitude state, we reset the oscillator to the low-amplitude state by a procedure described in Ref.~\onlinecite{Stambaugh2006b}. Figure~\ref{fig2}~(b) shows the histogram of the residence time, demonstrating an exponential decay of the state population. The linear fit determines the switching rate $W$. A similar behavior is observed for switching due to Gaussian noise\cite{Stambaugh2006b}. The measured switching rates for Poisson noise were close to the theoretical values that took the prefactor into account \cite{Billings2010}.

\begin{figure}[htbp]
\includegraphics[width=6.8cm]{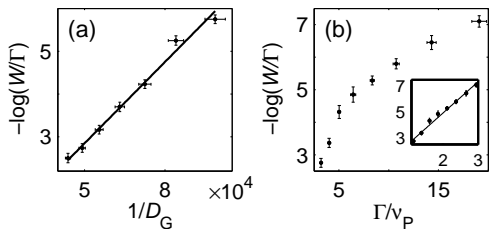}
\caption{(a) The logarithm of the switching rate scaled by the oscillator decay rate $-\log (W/\Gamma)$ for Gaussian noise as a function of the noise intensity $D_G$ for $\Delta\omega_{\eta}=3.14$~rad$~$s$^{-1}$; $D_G$ is in the units of s$^{-3}$.  The solid line is a linear fit. (b) The dependence of $-\log (W/\Gamma)$ on the reciprocal scaled pulse rate $\Gamma/\nu_P$ at $\Delta\omega_{\eta}=9.4$~rad$~$s$^{-1}$ for Poisson-noise induced switching; the pulse area $g_P$ is constant. Inset: the same data plotted vs.\ $\log(\Gamma/ \nu_P)$.}
\label{fig3}
\end{figure}

The dependence of the switching rate $W$ on the noise parameters and the distance to the bifurcation point is presented in Figs.~3-5. Figure~\ref{fig3}~(a) shows that, for Gaussian noise, $-\log W$ linearly depends on the reciprocal noise intensity $1/D_G$, in agreement with Eq.~(\ref{eq:QG_analytic}). The Poisson noise intensity $D_P = \nu_Pg_P{}^2/2$ depends on both the pulse rate $\nu_P$ and area $g_P$. In Fig.~\ref{fig3}~(b) we show that the increase of $-\log W$ with $1/\nu_P$ for fixed $g_P$ is strongly sub-linear. Replacing $1/\nu_P$ with $\log(1/\nu_P)$ gives a good linear fit, as seen from the inset in Fig.~\ref{fig3}~(b). This is completely different from the $1/D_G$ dependence for Gaussian noise and is consistent with Eq.~(\ref{eq:QP_analytic}).

\begin{figure}[htbp]
\includegraphics[width=6.8cm]{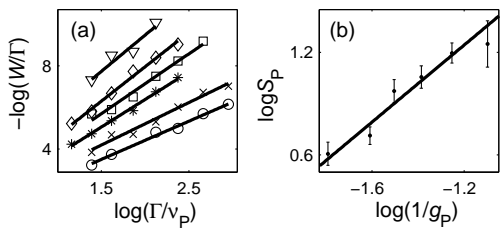}
\caption{(a)The logarithm of the scaled switching rate $-\log (W/\Gamma)$ vs. the logarithm of the reciprocal pulse rate  $\log (\Gamma/\nu_P)$ at $\Delta\omega_{\eta}= 9.4$~rad$~$s$^{-1}$. The solid lines are a linear fit for a fixed $g_P$. From top to bottom, the values of $g_P$ were $3, 3.5, 4, 4.5, 5, 6$, in scaled units. The error bars are smaller than the markers. (b) Logarithm of the slopes from (a) as a function of $\log(1/g_P)$. The line is a linear fit, yielding the exponent $\alpha =1.12\pm 0.14$.}
\label{fig4md}
\end{figure}

Next, the dependence of the switching rate on the pulse area $g_P$ was studied by changing the pulse height, while keeping the pulse width $t_P$ constant. In Fig.~\ref{fig4md}~(a) we show $-\log W$ as function of $\log(1/\nu_P)$ for different $g_P$. Fitting the data by straight lines, we found the slope $S_P=|d\log W/d\log\nu_P|\approx |d{\cal Q_P}/d\log \nu|$ as function of $g_P$. Figure~\ref{fig4md}~(b) shows a log-log plot of this slope. The measured exponent of $S_P\propto g_P{}^{\alpha}$ is $\alpha =1.12\pm 0.14$, consistent with the theoretical value of $1$ in Eq. (\ref{eq:QP_analytic}) [the factor $\log g_P$ in ${\cal Q}_P$, Eq.~(\ref{eq:QP_analytic}), could not be measured in the studied range of $g_P$].

Because the pulses of $f_P(t)$ ``push" the quadrature $Y(t)$ in one direction, near a bifurcation point, where the motion in the rotating frame is overdamped, switching is possible only where this is the direction from $Y_{\min}$ to $Y_{\max}$. For $g_P>0$, Eq.~(\ref{eq:QP_analytic}) describes switching near $\omega_{1B}$ in Fig.~\ref{fig1}. Switching near the second (larger-frequency, in our case) bifurcation point $\omega_{2B}$ occurs if $g_P<0$. This was tested in the experiment by changing sign of $f_P(t)$ in Eq.~(\ref{eq:Poisson_noise}). The switching exponent near $\omega_{2B}$ is given by Eq.~(\ref{eq:QP_analytic}) with $g_P\to |g_P|$ and $\eta\to |\eta|$.

\begin{figure}[htbp]
\includegraphics[width=6.8cm]{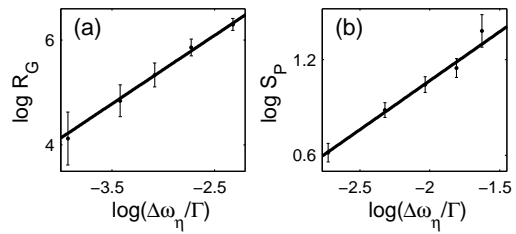}
\caption{The activation barrier $R_G$ for Gaussian noise induced switching (a) and the slope $S_P=|d\log W/d\log \nu_P|
$ for Poisson noise induced switching (b) as functions of the distance to the bifurcation point $\Delta\omega_{\eta}$. The fitted critical exponents are $1.31\pm0.20$ for Gaussian noise in panel (a) and $0.61\pm0.08$ for Poisson noise in panel (b).}
\label{fig5md}
\end{figure}

Last but not least, we studied the dependence of the switching exponent on the distance to the bifurcation point. For Gaussian noise, we generated a number of plots similar to Fig.~\ref{fig3}~(a), each at a different $\Delta\omega_{\eta}$.  We could then extract the activation barrier $R_G={\cal Q}_GD_G$ [the switching rate $W\propto \exp(-R_G/D_G)$] from the slope of ${\cal Q}_G$ vs $D_G^{-1}$. Figure~\ref{fig5md}~(a) shows $\log R_G$ vs $\log \Delta\omega_{\eta}$. A power law fit gives a critical exponent of $1.31 \pm 0.20$ in agreement with the theoretical predictions of $3/2$ in Eq.~(\ref{eq:QG_analytic}) and with other experiments \cite{Stambaugh2006b,Siddiqi2006}.

For Poisson pulses, the slopes $S_P$ were extracted from plots of $-\log W$ vs.  $\log(1/\nu_P)$ [similar to the inset of Fig.~\ref{fig3}~(b)] at different  $\Delta\omega_{\eta}$ and plotted as Fig.~\ref{fig5md}~(b). Fitting the results in Fig.~\ref{fig5md}~(b) with a power law $S_P\approx |d{\cal Q_P}/d\log \nu_P|\propto
\eta^{\xi}$ gives a critical exponent of $\xi =0.61 \pm 0.08$, consistent with the theoretical prediction of $1/2$ [Eq.~(\ref{eq:QP_analytic})].

Apart from the different exponent in the dependence on $\Delta\omega_\eta$ compared to the Gaussian noise, the dependence of ${\cal Q}_P$ on $\Delta\omega_{\eta}$ for Poisson pulses contains an extra log factor in Eq.~(\ref{eq:QP_analytic}).
The dynamical range of the parameters in our experiment did not allow us to single out this factor  on top of the $\Delta\omega_{\eta}^{1/2}$ directly by measuring $\log W$ for different $\Delta\omega_\eta$ while staying close to the bifurcation point. However, we made use of the fact that $\nu_P$ is only present inside the log factor in ${\cal Q}_P$. Therefore the slope $S_P$ should be independent of $\log\Delta\omega_{\eta}$. By varying the Poisson pulses via $\nu_P$, the factor $\Delta\omega_{\eta}^{1/2}$ could be isolated using the slope of the activation exponent in the inset of Fig.~\ref{fig3}~(b) as function of $\log\nu_P$ for different modulation frequency.

The parameters of the oscillator used in the experiment were chosen to be close to the bifurcation values. However, they were not close enough to make the oscillator motion in the rotating frame fully controlled by one dynamical variable $Y(\tau)$, i.e., the assumption that the variable $X(\tau)$ follows $Y(\tau)$ adiabatically was not strictly valid. In Appendix we show that the full theory of the escape rate in the studied range gives results very close to the asymptotic expression (\ref{eq:QP_analytic}). This is because $X(t)$ behaves nonadiabatically only very close to the metastable state of the oscillator, but the contribution of this area to the switching exponent ${\cal Q}_P$ is small.

\section{CONCLUSIONS}

Our results demonstrate that driven nonlinear systems can serve as bifurcational detectors of non-Gaussian noise, based on
the sensitivity to the noise statistics of the dependence of $\log W$ on the distance to the bifurcation point. The detectors can be implemented with nano-mechanical systems, rf Josephson junctions, or other systems. A stringent condition is that within the relaxation time of the system the mean number of pulses needs to remain small. As a rough estimate, to measure fluctuations in mesoscopic systems with current $\sim 1$~pA, the relaxation time of the detector needs to be $\sim 100$~ns. A nanomechanical detector will therefore require a state-of-the-art GHz resonator with a quality factor $\gtrsim 100$.

At the same time, non-Gaussian noise plays an increasingly important role in mechanical devices as they are getting smaller. Besides the discreteness of the fluctuating electric charge coupled to the mechanical motion \cite{Clerk2010}, it can come from attachment or detachment of molecules \cite{Jensen2008,Naik2009,Dykman2010} or random spin flips \cite{Rugar2004,Palyi2012}. The effects of non-Gaussian noise on nano-mechanical systems warrant further investigation.

In summary, we have measured the rate $W$ of Poisson noise induced switching in a nonlinear resonator near a bifurcation point where the number of stable vibrational states changes. The observed dependence of $\log W$ on pulse rate $\nu_P$ and area $g_P$, as well as on the distance to the bifurcation point $\eta$ is consistent with the predicted scaling \cite{Billings2010}, which is strongly different from the scaling for Gaussian-noise induced switching.

This work was supported by the NSF through grants No. CMMI-0856374 and 0900666. H.B.C. is supported by Shun Hing Solid State Clusters Lab and HKUST 600312 from the Research Grants Council of Hong Kong SAR.

\appendix

\section{THE MOST PROBABLE SWITCHING PATH}

The random trajectories followed by our periodically driven micromechanical resonator in fluctuation-induced switching  are concentrated around a specific path in its phase space $(X,Y)$, which for Gaussian noise has been now seen in experiment.\cite{Chan2008,*Chan2008c}  This path is commonly known as the most probable switching path (MPSP).  Even though the mechanism for switching induced by Poisson noise is qualitatively different from Gaussian noise as discussed in the main text, the notion of MPSP still applies.

In general, the MPSP follows complex patterns in multivariable systems. For example, in a micromechanical resonator driven into parametric resonance, if the motion in the rotating frame is underdamped the MPSP has been shown to consist of spirals around the two attractors on  the $(X,Y)$ plane. As the system approaches the bifurcation point, the motion near the corresponding stable state in the rotating frame undergoes a crossover from 2D to effectively 1D and becomes overdamped. In a switching event, the system most likely moves along a straight line on the $(X,Y)$ plane from the initially occupied state to the saddle point ; this motion is controlled by a 1D overdamped soft mode fluctuating in a 1D potential well. The scaling of the switching exponent predicted in Ref.~\onlinecite{Billings2010} is obtained using the asymptotic theory that considers the escape of the system from such potential well.

Here, with the parameters used in the experiment, we analyze theoretically the motion of our system in phase space. We show that under the experimental conditions, the portion of the MPSP that goes from the stable state to the saddle point (the ``uphill" portion) is largely linear. Spiraling only occurs close to the initially occupied attractor. For the spiraling to entirely disappear so that the motion is truly 1D, it is necessary to choose driving frequencies even closer to the bifurcation value. Nevertheless, the spiraling portion of the path contributes only a negligible amount to the action in the calculation of the switching exponent. As a result, the scaling behavior of the switching exponent is found to extend beyond the 1D regime, in agreement with our experimental findings.

\begin{figure}[h]
  \centering
  \begin{minipage}[b]{4 cm}
    \includegraphics[width = 4cm]{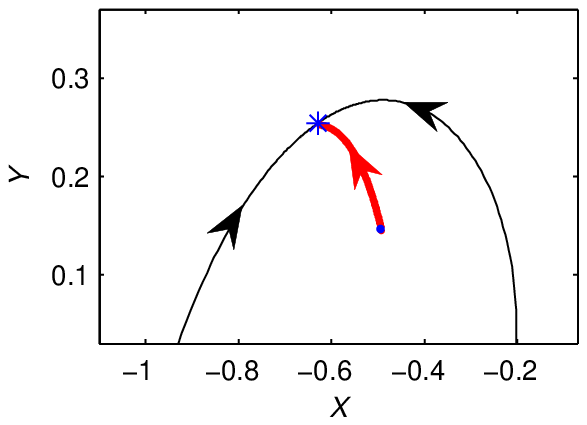}
  \end{minipage}
  \begin{minipage}[b]{4 cm}
    \includegraphics[width = 4cm]{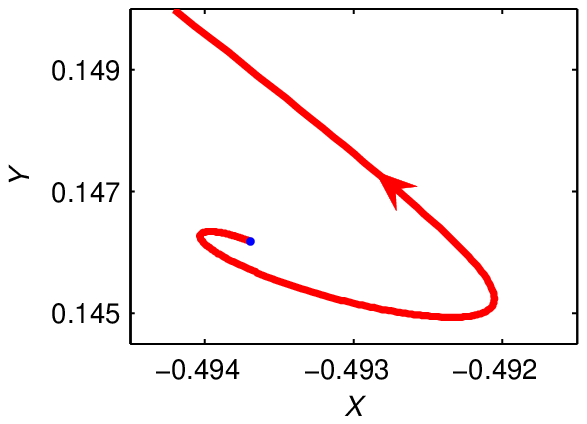}
  \end{minipage}
\caption{(Color online) (a) The uphill portion of the MPSP that is numerically calculated with the parameters $\Delta\omega_{\eta}=9.4$~rad s$^{-1}$,  $\nu_P=11.5$~Hz, and the dimensionless pulse area $\tilde g_P=3.5$. The MPSP is shown by the thick solid line (red) that goes from the attractor (dot) to the saddle point (star). On this scale, the spiraling of the MPSP is not resolved. The thin (black) lines are the portions of the separatrix which bounds the basin of attraction of the considered attractor. The other attractor (not shown) is far away from the saddle point. (b) Close-up of the MPSP near the attractor. A spiraling behavior can be observed. }
\label{fig:Traj}
\end{figure}

In the rotating frame, the generalized Fokker-Planck equation for the probability density $\rho$ of the system in the presence of Poisson noise reads\cite{Billings2010}
\begin{equation}
\dot\rho=-\partial_{\mathbf{q}}[\mathbf{K}\rho(\mathbf{q})]+\tilde\nu_P[\rho(\mathbf{q}-\tilde{\mathbf{g}}_P)-\rho(\mathbf{q})],
\label{eq:FP2D}
\end{equation}
where $\mathbf{q}=(X, Y)$ is the position vector in the rotating frame, $\mathbf{K}=(K_X, K_Y)$, and $\tilde{\bf g}_P=(0,\tilde g_P)$. The components $K_{X,Y}$ are
given by Eq.~(\ref{eq:eom_full}); $\tilde g_P$ and $\tilde\nu_P$ are the dimensionless pulse area and pulse rate, respectively.

We are interested in the quasistationary solution of Eq.~(\ref{eq:FP2D}), which is formed over the relaxation time and persists for times small compared to the reciprocal  rate of switching from the considered metastable state. Respectively, we seek the solution as $\rho(\mathbf{q})=\exp[-s(\mathbf{q})]$.

We assume that on the tail of the distribution $s({\bf q})$ is large, but that it is smooth on the scale $\sim |\tilde{\bf g}_P|$. Then, if we keep terms $\propto \partial_{\bf q}s$ and assume them large, but disregard terms with higher derivatives of $s$, we can reduce Eq.~(\ref{eq:FP2D}) to equation
\begin{equation}
\label{eq:HamiltonianFP}
H({\bf q},{\bf p})=0,\qquad H=\tilde\nu_P[\exp(\mathbf{p}\tilde{\bf g}_P)-1]+\mathbf{p}\cdot\mathbf{K},
\end{equation}
where ${\bf p}\equiv \partial_{\bf q}s$. This equation maps the problem of noise-induced switching on the problem of Hamiltonian dynamics of an auxiliary system. This system has the coordinate vector ${\bf q}$ and the momentum vector $\mathbf{p}$, its Hamiltonian function is $H({\bf q},{\bf p})$, whereas function $s$ is its action.\cite{Jordan2005,Billings2010}  The uphill portion of the MPSP is a heteroclinic orbit that connects the fixed points  $(\mathbf{q}_a,{\bf p}_A={\bf 0})$ and $(\mathbf{q}_{\cal S},{\bf p}_{\cal S}={\bf  0})$, where ${\bf q}_A$ and ${\bf q}_{\cal S}$ correspond to the positions of the attractor and the saddle point of the resonator in the rotating frame.

The switching exponent ${\cal Q}_P$ is given by the integral of $\mathbf{p}$ along the MPSP,
\begin{equation}
{\cal Q}_P=\int\nolimits_{\mathbf{q}_A}^{\mathbf{q}_{\cal S}}\mathbf{p}\,d\mathbf{q}.
\label{eq:QPnumerical}
\end{equation}

\begin{figure}
  \centering
  \includegraphics[width = 8.5cm]{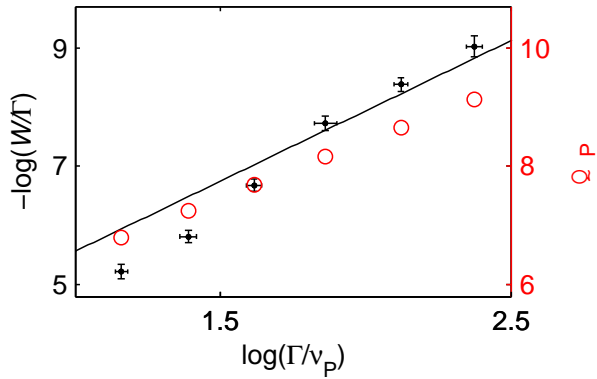}
\caption{(Color online) The measured values of $-\log (W/\Gamma)$ (small solid circles) agrees with the asymptotic theory (the black line) for typical experimental parameters: $\Delta\omega_{\eta}=9.4$~rad s$^{-1}$ and $\tilde g_P=3.5$ in scaled unit. The numerical calculated ${\cal Q}_P$ (red circles) are plotted with the right vertical axis that is shifted by $1$ compared to the left one. }
\label{Fig:ComparisonQP}
\end{figure}

We have numerically calculated the MPSP taking into account that close, but not too close to the fixed point $({\bf q}_A,{\bf 0})$, action $s({\bf q})$ is quadratic in ${\bf q}-{\bf q}_A$. \cite{Dykman2010a} Similar to how it is done for Gaussian noise (cf. Ref.~\onlinecite{Chan2008} and papers cited therein), one can find MPSP using the shooting method by finding the ``right" direction for the Hamiltonian trajectories of the auxiliary system that go from the vicinity of $({\bf q}_A,{\bf 0})$, so that they approach $({\bf q}_{\cal S},{\bf 0})$.

Figure~\ref{fig:Traj} shows the uphill portion of a MPSP for typical values of the experimental parameters.  The trajectory does not show any spiraling at this length scale. A close-up of the region close to the attractor [Fig.~\ref{fig:Traj} (b)] reveals that spiraling indeed occurs. Therefore the uphill path is, straightly speaking, not 1D. However, since the spiraling portion is small and lies in the region of small $|{\bf p}|$, whereas much of the uphill path remains largely straight, the asymptotic theory provides a good estimation of the switching rate.

In Fig.~\ref{Fig:ComparisonQP}, we compare the measured $-\log(W/\Gamma)$ with the theoretical values calculated from the the asymptotic theory for 1D dynamics [Eq.~(\ref{eq:QP_analytic})] and the numerical values that go beyond the 1D theory. The black line represents the results for $-\log(W/\Gamma) ={\cal Q}_P-\log({\cal C}/\Gamma)$, where ${\cal Q}_P$ and the prefactor $\mathcal{C}$ are both calculated from the asymptotic 1D theory.\cite{Dykman2010a} The good agreement indicates that the asymptotic theory provides a good estimate of the switching exponent even beyond the strictly 1D regime.

The numerical results for the switching exponent ${\cal Q}_P$ [Eq. (\ref{eq:QPnumerical})] refer to the same parameters. Since the numerical method does not yield the prefactor $\mathcal{C}$, it is not possible to perform a direct comparison between the numerically calculated ${\cal Q}_P$ and the experimentally measured $-\log W/\Gamma$. In Fig.~\ref{Fig:ComparisonQP}, , the right vertical axis (corresponding to the numerically found  ${\cal Q}_P$) is shifted from the left vertical axis (corresponding to $-\log W/\Gamma$). The plot indicates that the slope of ${\cal Q}_P$ vs. $\log \Gamma/\nu$ is in good agreement with the results from both the asymptotic theory and the measurement.

\bibliographystyle{apsrev4-1}

%

\end{document}